\documentclass[useAMS,usenatbib,usegraphicx]{mn2e}
\usepackage{amssymb}
 
\title[Massive Galaxies at $1<z<2$]{Gemini K-band NIRI Adaptive Optics Observations of
Massive Galaxies at $1<z<2$}
\author[Carrasco, Conselice \& Trujillo]{Eleazar R. Carrasco$^{1}$\thanks{E-mail:
rcarrasco@gemini.edu}, Christopher J. Conselice$^{2}$, Ignacio Trujillo$^{3,4}$\\
$^{1}$Gemini Observatory/AURA, Southern Operations Center, Casilla 603, La Serena, Chile \\
$^{2}$University of Nottingham, School of Physics \& Astronomy, Nottingham, NG7 2RD UK  \\
$^{3}$Instituto de Astrof\'isica de Canarias, E-38205, La Laguna, Tenerife, Spain; Ram\'on y Cajal Fellow\\ 
$^{4}$Departamento de Astrof\'isica, Universidad de La Laguna, E-38205 La Laguna, Tenerife, Spain}

\def\solm{M$_{\odot}\,$}

\def\solm{M$_{\odot}\,$}

\def\mass{$10^{11}$ M$_{\odot}\,$}

\def\casgm20{CAS-G-M$_{20}\,$}
\def\m20{M$_{20}\,$}
\begin{document}

\date{Accepted 2010 March 07; Received 2010 March 01; in original form 2010 January 27}
\pagerange{\pageref{firstpage}--\pageref{lastpage}} \pubyear{2010}

\maketitle

\label{firstpage}

\begin{abstract}

We present deep K-band adaptive-optics observations of eight very massive
(M$_{*}\sim4\times10^{11}$ \solm) galaxies at $1 < z < 2$ utilizing the Gemini NIRI/Altair Laser Guide 
System. These systems are selected from the Palomar Observatory Wide-Field Infrared (POWIR) survey, 
and are amongst the most massive field galaxies at these epochs. The depth and high spatial resolution of 
our images allow us to explore for the first time the stellar mass surface density distribution of massive
distant galaxies from 1 to 15 kpc on an individual galaxy basis, rather than on stacked images.  We confirm 
that some of these massive objects are  extremely compact with  measured effective radii between 
0\farcs1 - 0\farcs2, giving sizes which are $\lesssim 2$ kpc, a factor of $\sim$7 smaller in effective 
radii than similar  mass galaxies today. Examining stellar mass surface densities  as a function of fixed physical
aperture, we find an over-density of material within the inner profiles, and an under-density in the outer 
profile, within these high-$z$ galaxies compared with similar mass galaxies in the local universe.
Consequently, massive galaxies should  evolve in a way to decrease the stellar mass density in their
inner region, and at the same time creating more extensive outer light envelopes. We furthermore show that 
$\sim$38\%$\pm20$\% of our sample contains evidence for a disturbed outer stellar matter distribution 
suggesting that these galaxies are undergoing a recent dynamical episode, such as a merger or accretion event. 
We calculate that massive galaxies at $z < 2$ will undergo on the order of five of these events, a much higher
rate than observed for major mergers, suggesting that these galaxies are growing in size and stellar 
mass in part through minor mergers during this epoch.  

\end{abstract}

\begin{keywords}
galaxies:  evolution - galaxies: formation - galaxies: structure - galaxies: high redshift
\end{keywords}

\section{Introduction}

One of the most remarkable findings in extragalactic astronomy during the past few years is that the
most massive galaxies in the universe at $z>$1, were much smaller in the past than galaxies of the same
stellar mass are today (Daddi et al. 2005; Trujillo et al. 2006; 2007; Longhetti et al. 2007; Toft et
al. 2007; Zirm et al. 2007; Buitrago et al. 2008; Cimatti et al. 2008; van der Wel et al. 2008;
Stockton et al. 2008; van Dokkum et al. 2008; Damjanov et al. 2008; Cassata et al. 2009; Mancini et al.
2010).  The effective radii of galaxies with stellar masses M$_{*}>$ \mass is observed to be, on average,
a factor of $\sim$ four smaller at $z \sim 2$ compared to galaxies of the same mass today (e.g., Trujillo et al. 2006,
2007;  Buitrago et al. 2008).  The origin of this apparent evolution is not clear, nor is it certain
that this effect is not partially, or entirely, due to redshift effects that make distant galaxies look
more compact, or less massive, than they actually are due to the  difficulty of measuring accurate 
masses and sizes at higher redshifts (see e.g. a discussion in Hopkins et al. 2010). However, despite these
uncertainties, it is unlikely that the small sizes of high redshift galaxies can be entirely explained
by observational uncertainties, given the rapidly increasing number of observations utilizing different depths and
instruments, and the confirmation that these objects are truly massive through the first measurements
of their stellar kinematics (Cenarro \& Trujillo 2009; van Dokkum, Kriek \& Franx 2009; Cappellari et
al. 2009).

On the other hand, if distant massive galaxies are more compact in the past  than they are today,
then an important question remains concerning how  these galaxies grew in size over several Gyr. 
This problem can be put  in context in a number of ways, including: (1) the number and mass
densities of massive galaxies at $z<1$ do not evolve much more than about a factor of two (e.g.,
Conselice et al. 2007), and (2) there are basically no compact massive galaxies today
(Trujillo et al. 2009; Valentinuzzi et al. 2009; Taylor et al. 2009).  Thus it is not easy to
simply increase the masses of all galaxies through significant merging or growth through star formation.  Nor
would this explain why the relation between size and stellar mass has evolved so significantly over
the history of the universe.  It remains possible that major and/or minor mergers (Khochfar \& Silk
2006; Hopkins et al. 2009; Naab et al. 2007; Naab, Johanson \& Ostriker 2009; Nipoti et al. 2009),
star formation or the action of AGN (Fan et al. 2008) all potentially increase the measured
effective radii of massive galaxies, yet the solution to this problem thus far remains unknown.

To shed light on this issue we have obtained very deep and high resolution K-band  imaging of 
a small sample of eight massive galaxies in the Extended Groth Strip (EGS) for which we can 
study these galaxies in the rest-frame optical (around I-band rest-frame) to better compare 
with lower redshift systems, as well as to minimize the effects of star formation that can 
potentially influence the measurement of radii. To the best of our knowledge this is 
the largest sample of massive galaxies at  $z>$1 observed in the K-band through adaptive
optics (AO). Previous studies have explored a smaller number of galaxies using the 
NIRC2 camera at Keck (three objects in the H-band; van Dokkum et al. 2008 and two objects in the 
J-band; Stockton et al. 2008). Small samples of these massive high-redshift galaxies at z$\sim$1.5 have 
been also imaged with the NICMOS camera (e.g. Trujillo et al. 2007; McGrath et al. 2008). The resolution 
of our images (PSF FWHM  $\lesssim$ 0\farcs15 or equivalently $\sim$1.3 kpc at z=1.5) is around two times
better than observations taken with NICMOS ($\sim$ 0\farcs3) and marginally 
better than those presently achievable with the HST WFC3 ($\sim$0\farcs18; Cassata et el. 2009).

It has been recently claimed that these massive compact galaxies are growing inside-out, with
a central core forming early-on, and the rest of the mass gradually accumulating in the outer
regions. This suggestion is based on the analysis of surface brightness profiles created by stacking a
large number of galaxies using a poorer resolution than what we use in this paper (Hopkins et al.
2010; Bezanson et al. 2009; van Dokkum et al. 2010).  We present very deep K-band imaging 
to explore, on individual basis, the above hypothesis. Our surface brightness profiles
reach $\sim$25.5 mag/arcsec$^2$ (AB system) in the observed K-band, the deepest of such observations
yet  presented in the near-infrared.

In addition, the quality of our data allow us to explore the existence of nearby fainter companions, as
well as possible mergers signatures in our galaxies that can possibly shed light on the main evolutionary paths
followed by massive galaxies in order to increase their sizes during the last $\sim$10 Gyrs. We find
that a substantial fraction of our galaxies show evidence for tidal distortions and interactions with
other galaxies. While this observation by itself does not prove that these systems are becoming larger
due to mergers, it does suggest that this process is at least partially responsible for 
increasing the sizes of massive galaxies over time.

This paper is organized as follows: \S 2 describes the new data
and observations, \S 3 gives our analysis of the data and
some conclusions, while \S 4 is a summary of our work.
We use a standard cosmology of H$_{0} = 70$ km s$^{-1}$ Mpc$^{-1}$, and 
$\Omega_{\rm m} = 1 - \Omega_{\lambda}$ = 0.3 throughout.

\section{Sample and Data}

The eight galaxies within our sample were taken from the Palomar Observatory
Wide-Field Infrared (POWIR) survey (Conselice et al. 2007; 2008).   The larger 
sample of massive galaxies in which our current galaxy sample was taken is presented 
in Conselice et al. (2007).  This sample is selected based on having a large stellar mass,
with M$_{*} > 10^{11}$ \solm, and to be at redshifts 1$<$z$<$2. Our stellar masses are 
photometrically determined based on optical and NIR photometry (see Conselice et al. 2007).
The sample we use in this paper contain spectroscopic redshifts measured through the 
DEEP2 spectroscopic survey, or through photometric redshifts calculated through optical and
near infrared photometry (see Conselice et al. 2007 for the details behind this).  The three systems 
at $z < 1.4$ have spectroscopic redshifts, and the remaining systems have photo-zs.  The photo-z 
accuracy for this sample of massive galaxies is very good, with an accuracy of $\delta z/(1+z) = 0.025$ 
(Conselice et al. 2007).

\begin{figure*}
\includegraphics[width=17.5cm]{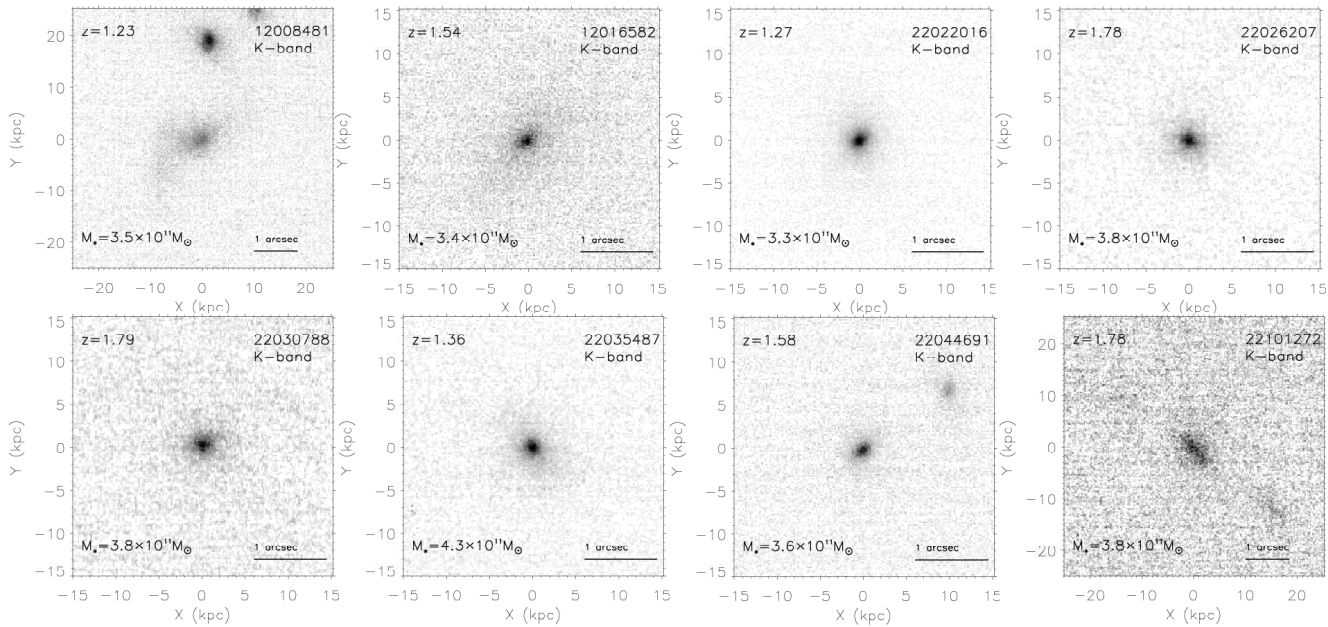}
\caption{K-band Gemini observations of our sample of eight galaxies.  Listed on each is the galaxy's name,
which band the image is taken in (always K), the stellar mass and the redshift. The solid line
indicates 1 arcsec angular size.}
\label{newfig}
\end{figure*}

The new observations presented in this paper were obtained with the Gemini-North telescope using the 
NIRI  (Hodapp et al. 2003) with the Altair/LGS (laser guide star) adaptive optics systems  (Herriot et 
al. 2000, Boccas et al. 2006).  Utilizing this system requires that a bright star ($\lesssim$ 18 mag in R-band)
be positioned near the object of interest.  Therefore, our sample of massive galaxies are those which are 
within 15\arcsec\, of a relatively brighter star to facilitate a maximum correction during the observations. 
We searched through the POWIR archive to obtain a sample of massive galaxies at $1 < z < 2$ which are near 
enough to a star brighter than $R=18$ mag  to facilitate these observations with NIRI and Altair/LGS.  
The final sample of massive galaxies has AO tip-tilt stars with magnitudes between 15.9 and 17.6 mag in R-band
located within 15\arcsec\, from the galaxy of interest.

The galaxies were imaged through the K (2.2 $\mu$m) filter in the Spring and Autumn of 2008 and 2009, 
in queue mode. We used the f/32 camera which provided a field of view of 22\farcs5 $\times$ 22\farcs5 
using the 1024 $\times$ 1024 ALLADIN InSb detector, with a pixel scale of 0\farcs0219 on 
side. Between 63 to 72 images of 60 s per galaxy were obtained, given effective exposure times of  3780 sec 
to 4320 sec. In addition, photometric standard stars were observed before and after our galaxies. By monitoring 
photometric standard stars and the field stars throughout our observations, we find that 
the effective seeing throughout was between 0\farcs08 and 0\farcs15, varying with stellar 
magnitude, and the location of the star relative to the galaxy of interest. 

We reduced our data by using standard procedures for near-infrared imaging provided by the Gemini Observatory 
through the NIRI package inside IRAF.  The reductions steps applied to the data are: (1) the normalized flats 
were constructed from flat images observed with the calibration unit with the shutter closed 
(lamps off) and with the shutter open (lamps on).  Flat images observed with the shutter 
closed were also used to identify bad pixels; (2) the sky images were constructed 
from the raw science images by identifying objects in each frame, removing 
them, and averaging the remaining good pixels (the images were observed with a 
dither offsets of 2\arcsec~- 3\arcsec, depending on the galaxy); (3) the raw 
science images were processed by subtracting the sky on a frame-by-frame basis and
dividing by the normalized flat field images; (4) The final processed images were then
registered to a common pixel position and median combined. The field of view
of the final images are between 16\farcs5 $\times$ 16\farcs5 and 18\farcs5
$\times$  18\farcs5.   The images for our eight galaxies are shown in Figure~1.

\section{Analysis}

The structural parameters presented in this paper were estimated using the GALFIT code (Peng et al.
2002). Sizes are parameterized by the half-light or effective radius along the semi-major axis a$_e$, and
are measured and then circularized, r$_{e,circ}$=a$_e$$\sqrt{b/a}$, with b/a being the axis ratio of
the isophotes describing the object.

GALFIT convolves model S\'ersic (1968) r$^{1/n}$  galaxy models with the PSF of the images and determines
the best fit by comparing the convolved model with the observed galaxy surface brightness distribution using a
Levenberg-Marquardt algorithm to minimize the $\chi^2$ of the fit. The S\'ersic model is a flexible
parametric description of the surface brightness distribution of galaxies and contains the
exponential ($n=1$) and de Vaucouleurs ($n= 4$) models as particular cases. In addition, this model is
used in the structural analysis of the SDSS galaxy sample (our local comparison sample; Blanton et al.
2003; Shen et al. 2003).

The S\'ersic index $n$ measures the shape of surface brightness profiles. In the nearby Universe,
galaxies with  n$<$2.5  are mostly disc-like objects, whereas galaxies with  n$>$2.5  are mainly
spheroids (Ravindranath et al. 2002). We use this S\'ersic index criterion to split our sample at
higher redshifts and to facilitate a comparison with the local galaxy population. During the fit,
neighboring objects are fitted simultaneously. The quantitative results of our fitting are 
shown in Table 1.

To explore this change we have compared the differences between the PSF obtained from the 
AO tip-tilt star used in photometric standard fields observed before and after the galaxies and, 
when possible, the stellar PSFs found inside some of our fields. We find that that the tip-tilt stars 
in the photometric standard fields have a FWHM of $\sim$0.08\arcsec whereas the stars found 
inside our fields have $\sim$0.15\arcsec.

As we are analyzing images obtained using adaptive optics it is necessary to conduct an analysis of 
the robustness of the structural parameters to changes in the Strehl ratio through the images. 
To explore this change we have compared the differences between the PSF obtained from the 
AO tip-tilt star used in photometric standard fields observed before and after the galaxies, and 
the PSFs found inside some of our fields. We find that that the tip-tilt stars 
in the photometric standard fields have a FWHM of $\sim$0\farcs08 whereas the stars found 
inside our fields have $\sim$0\farcs15. To explore the effect of this degradation has in our 
structural parameters we have created a set of PSFs with different FHWMs: 0\farcs15 and 
0\farcs20 (this latter PSF is to account for extremely bad degradation cases). To build these mock 
PSFs we take the original AO tip-tilt star and convolve this PSF with a Gaussian to match the 
FWHM we are interested in. We compare these mock PSFs with those found in some of our fields 
and find that the agreement is very good. We then use GALFIT to fit our galaxies with the AO PSF and 
with the two new mock PSFs.  We find a slight change in the effective radii of our objects at the level 
of $<$15\%  using these different PSFs. We use this uncertainty as the typical error in our estimate 
of the sizes of our galaxies. In the case of the S\'ersic index $n$ we find two different behaviors, 
depending on whether the object is disk-like ($n \sim$1), or spheroid-like ($n \sim$4). For objects 
with $n \sim$ 1 the uncertainty in the index $n$ is $\sim$20\%,  whereas for objects with $n \sim$4 
the change can be as large as 50\%. We  use these two values as the typical uncertainty for this 
parameter. In Table~1, we list these values as measured using the PSF with FHWM equal to 0\farcs15.

We have also investigated our ability to retrieve the size and $n$ index values based on  
simulations (e.g. Trujillo et al. 2006; 2007) using mock galaxies that share magnitudes and effective 
radii as our sample galaxies. Based on these simulations, which are created to explore imaging at a
poorer  resolution and lower signal-to-noise, we find that the depth and resolution of our present data 
are more than sufficient to retrieve accurately our structural parameters. In fact, the above 
simulations suggest that, if anything, our size estimations are upper limits.

To obtain stellar mass density profiles, which we show in the next subsection, we measure 
directly the surface bright- ness profiles in the observed K-band for each of our galaxies. 
These profiles were extracted using circular apertures in an iterative way to guarantee a 
proper sky subtraction. The flux in different circular apertures is then extracted well beyond 
where the galaxy is plainly visible. By plotting fluxes at these apertures as a function of radius 
we determine at which radial distance the ellipses are no longer tracing the galaxy, and essentially
becomes a measure of the background, flat flux, or noise. We measure the mean value and 
standard deviation of the fluxes in this flat region. We then use this estimate of the sky to 
remove this component from our images, and we then extract the profile again in circular apertures. 
To specify the error on the profiles from our sky measurements we define a critical surface brightness 
to where we trust the final profile. This is the location where the surface brightness profiles 
obtained by either over- or under- subtracting the sky by $\pm$1$\sigma$ begin to deviate 
by more than 0.2 mag. The typical value for the full sample of this critical surface brightness is 
$\mu_{crit}$$\sim$25.5  K-mag/\arcsec$^2$. The mass profiles we show in the next section 
are at a limit brighter than this.

\begin{table*}
 \centering
 
  \caption{Properties of the Gemini high resolution sample galaxies}

  \begin{tabular}{ccccccccccc}
  \hline
Galaxy & R.A.   & Dec.   & a$_e$     & n & b/a & r$_{e,circ}$ & K$_{mag}$ & z & Log M$_{*}$     & r$_{e,circ}$  \\
  ID   & (2000)  & (2000) & (\arcsec) &   &     &   (\arcsec)   &   AB      &   &  (M$_{\sun}$) &  (kpc) \\

\hline

12008481 & 214.291901 & 52.477417 & 0.77$\pm$0.12 &  1.47$\pm$0.29 & 0.45 & 0.52$\pm$0.08 & 19.65 &  1.23 & 11.55 & 4.32$\pm$0.65 \\
12016582 & 214.408249 & 52.600197 & 1.11$\pm$0.17 &  3.97$\pm$1.99 & 0.56 & 0.83$\pm$0.12 & 20.37 &  1.54 & 11.54 & 7.03$\pm$1.05 \\
22022016 & 252.747849 & 34.891380 & 0.60$\pm$0.09 &  6.05$\pm$3.03 & 0.73 & 0.51$\pm$0.08 & 19.54 &  1.27 & 11.53 & 4.26$\pm$0.64 \\
22026207 & 253.076126 & 34.902084 & 0.23$\pm$0.03 &  2.54$\pm$1.22 & 0.83 & 0.21$\pm$0.03 & 20.59 &  1.78 & 11.59 & 1.77$\pm$0.27 \\
22030788 & 252.568848 & 34.911388 & 0.24$\pm$0.04 &  3.31$\pm$1.66 & 0.79 & 0.21$\pm$0.03 & 20.72 &  1.79 & 11.59 & 1.77$\pm$0.27 \\
22035487 & 252.908646 & 34.986378 & 0.65$\pm$0.10 &  4.61$\pm$2.30 & 0.71 & 0.55$\pm$0.08 & 20.06 &  1.36 & 11.64 & 4.63$\pm$0.69 \\
22044691 & 252.797272 & 35.033882 & 0.18$\pm$0.03 &  3.59$\pm$1.80 & 0.51 & 0.13$\pm$0.02 & 20.27 &  1.58 & 11.56 & 1.01$\pm$0.15 \\
22101272 & 253.146851 & 35.088760 & 0.45$\pm$0.07 &  1.02$\pm$0.20 & 0.52 & 0.33$\pm$0.05 & 20.73 &  1.78 & 11.58 & 2.79$\pm$0.42\\

\hline
\label{data}
\end{tabular}
\end{table*}

\subsection{Sizes and Densities}

The distribution of size and stellar mass for our objects is shown in Fig. 2. Our galaxies fall in the
same region as the observed $z>1$ size vs. stellar mass relation obtained using HST ACS imaging 
(Trujillo et al. 2007). As was mentioned in the Introduction,  a major criticism of the finding that 
distant galaxies are more compact than nearby ones is that perhaps the outer portions of the light 
from these distant galaxies is missed in the observations.  However, our Gemini data are very deep 
and we are able to probe to large radii within our observations.  In fact, our images are so deep 
that we can detect envelopes if they are actually present. For example, for two of our galaxies 
that have similar magnitude, S\'ersic index and redshift (i.e. 12016582 and 22044691) we can see 
in one of them  (12016582) an envelope (and its size is consequently large and falls in within the 
local relation)  whereas the other galaxy (22044691)  does not show any outer envelope 
(confirming its compactness). These results show that 
some of our galaxies are truly compact, and that we are not missing outer light that would increase their
measured sizes.

While our fitted sizes suggest that some of these galaxies are quite compact, we furthermore examine the 
stellar mass densities of our galaxies as a function of fixed physical aperture. We have plotted our 
individual galaxy stellar mass surface density profiles in Fig. 3. These stellar mass density profiles 
have been measured as follows. We have used the circularized observed surface brightness distribution
in the K-band images and the stellar mass estimates (obtained from the total flux within an aperture 
of radius 2$\arcsec$, equivalent to $\sim$17 kpc at these redshifts; Conselice et al. 2007) of 
these galaxies. We have assumed that galaxies have a negligible color gradient, and consequently that 
they can be characterized by a single stellar M/L ratio. As shown in Fig. 3, the depth and high resolution
of our images allow us to explore for the first time on an individual galaxy basis the stellar mass
density profiles of these very massive galaxies at high$-z$ from 1 kpc to 15 kpc. This means that we can
typically explore the stellar mass distribution of our galaxies up to 5r$_e$ in radius. 
This is equivalent to moderately deep observations of modern massive elliptical 
galaxies  (Caon et al. 1994).

\begin{figure*}
\includegraphics[angle=0, width=160mm]{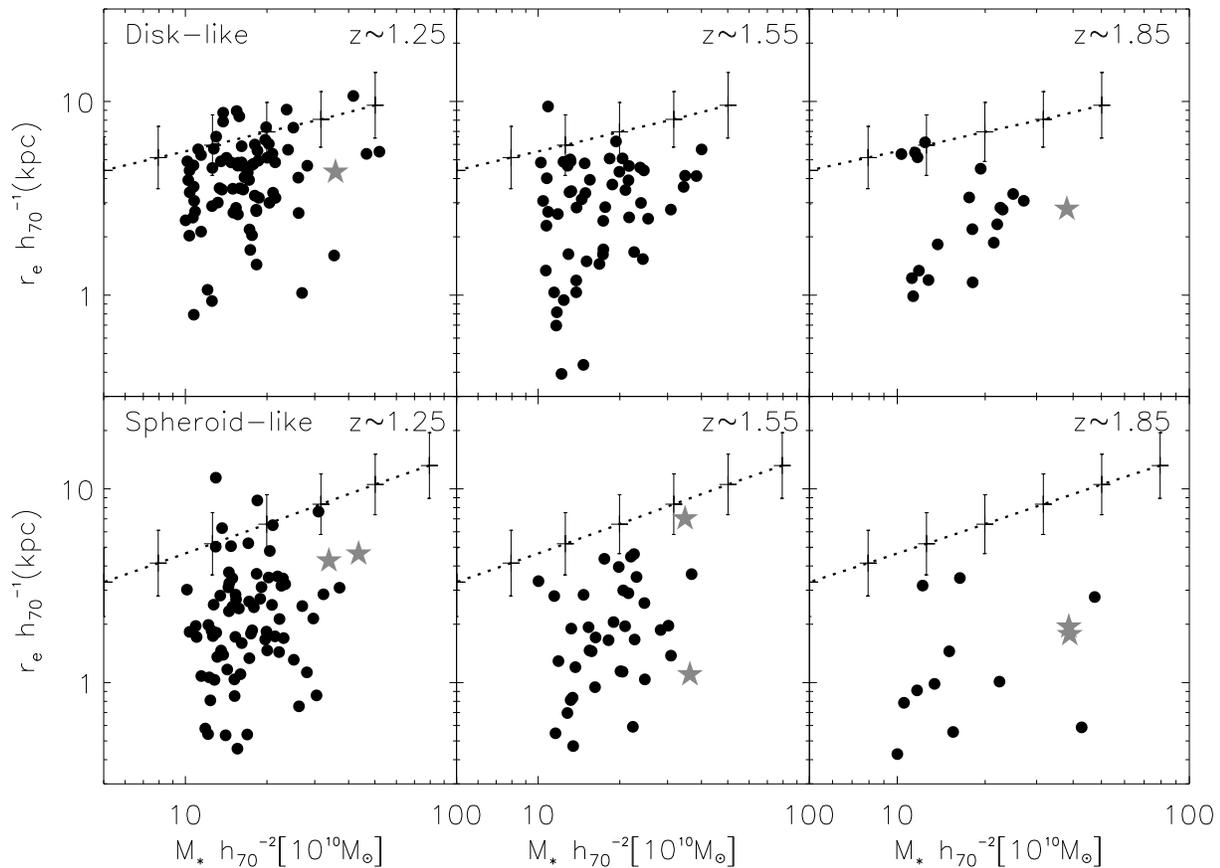}
 \caption{
The size-stellar mass relation for our sample galaxies (grey stars) over-plotted on points from ACS imaging 
of massive galaxies within the EGS sample (solid circles) as published in Trujillo et al. (2007). The dash-line 
is the mean and dispersion of the observed half-light radii of the SDSS late-type  ($n<$2.5) and early-type
($n>2.5$) galaxies as a function of stellar mass. We use the SDSS sample as the local reference  ($z \sim 0.1$) 
to compare with the galaxies in our high-z sample. The SDSS sizes were determined using a circularized 
S\'ersic model, and stellar masses were measured using a Kroupa IMF. SDSS sizes were measured using 
the observed r' band, which closely matches the V-band rest-frame filter at  $z \sim 0.1$.}
\label{allgalaxiesevol}
\end{figure*}

\begin{figure*} 
\includegraphics[angle=0, width=160mm]{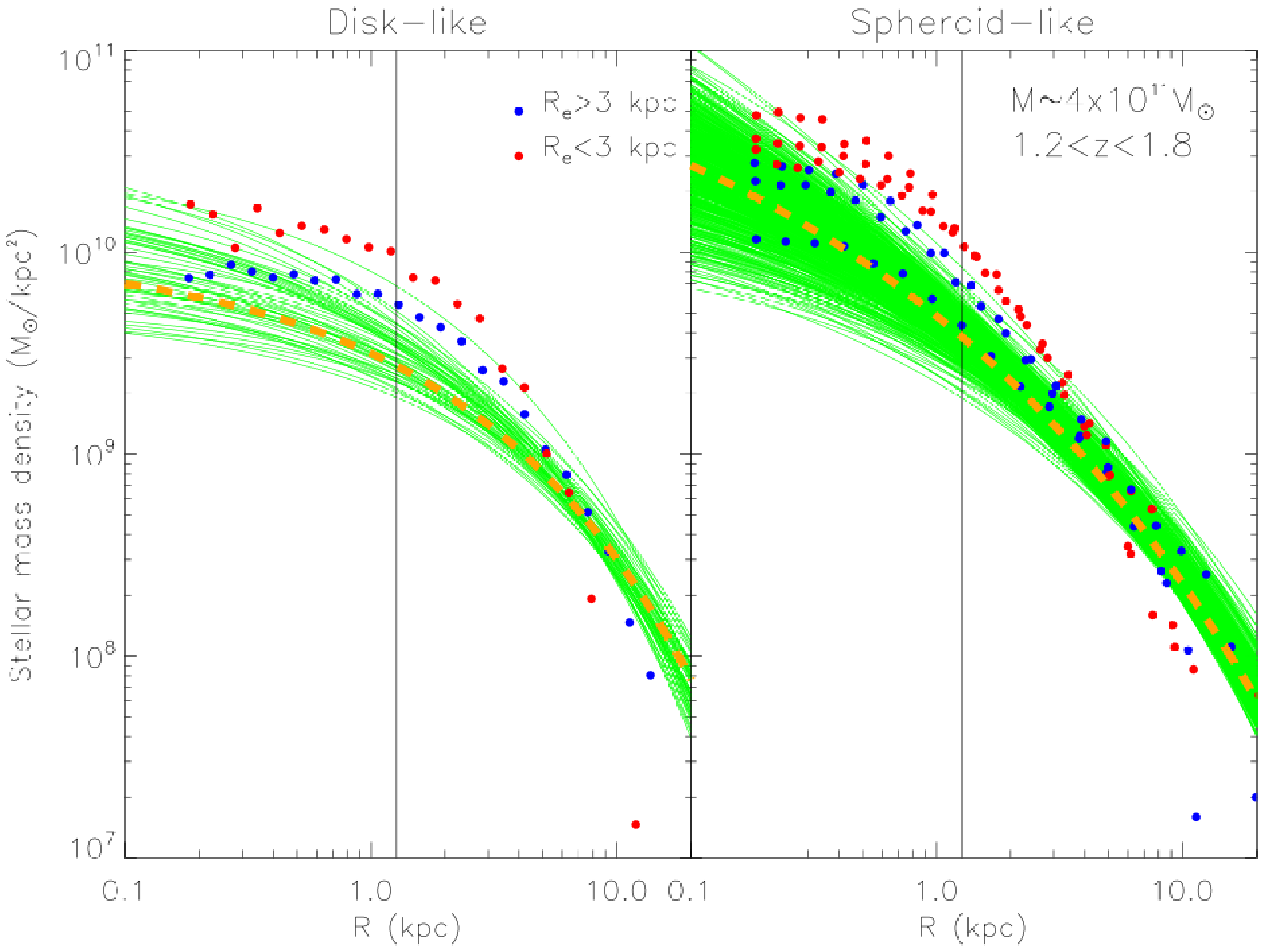}
\caption{Stellar surface mass density profiles. Our sample is split according to the S\'ersic 
index value. Those objects with a disk-like morphology (i.e. n$<$2.5) are shown in the left  panel, 
whereas galaxies with spheroid-like morphology (i.e. n$>$2.5) are shown in the right  panel. In 
addition, we have colored differently our galaxies depending on whether they are compact or not 
using the following criteria: r$_e>$3 kpc objects (blue) and r$_e<$3 kpc objects (red).
In addition, we show with green lines the stellar surface mass density profiles of nearby objects with
M$_{*}$$>$4$\times$10$^{11}$ \solm obtained from the SDSS DR7. The orange  dashed-lines show the median 
stellar surface mass density profile for these nearby galaxies.  The vertical line shows the equivalent 
size in kpc of a FWHM PSF of 0\farcs15 at $z = 1.5$. The depth and high resolution of our images 
allows us to explore the profiles of our sample galaxies from 1 to 15 kpc, reaching several times 
the effective radii of these objects.} 
\label{allmassprofile} 
\end{figure*}

To put our galaxies in context we compare our high$-z$ galaxy stellar mass density profiles with the 
profiles of  present day galaxies with equal, or larger, stellar masses than our sample. Our local 
comparison sample is obtained from the SDSS DR7, based on the NYU value  added catalog from Blanton 
et al. (2005).  This provides a list of galaxies as a function of redshift, stellar mass 
and surface brightness fitted parameters, such as the Sersic index $n$.  The exact selection
of these comparison objects are: redshift $0 < z < 0.2$, stellar  masses (using Kroupa IMF, which 
nearly matches our Chabrier IMF) at M$_{*}$$>$4$\times$10$^{11}$ \solm. Spheroid like galaxies  
are selected using $n>2.5$.  There are a total of 782 such galaxies with a median S\'ersic 
index of $n=4.0$, and a median size of r$_{\rm e}=14.7$ kpc. Disk like galaxies from the SDSS are 
selected using $n<2.5$, and there are a total of 72 objects with median S\'ersic index $n=2.12$,
and a median radius r$_{\rm e}=12.4$ kpc.

As Figure~3 shows, our high-z galaxies have higher stellar mass densities near their centers, 
and  lower densities in their outer regions compared to today's massive galaxies. Interestingly, 
galaxies with larger sizes have stellar mass surface densities which approach the observed mass 
distribution found locally in the SDSS data.  We have quantified the different mean stellar surface 
mass densities at different physical radii for our compact, and non-compact, galaxies compared to the 
local sample. The results of our analysis are presented in Table 2. As a comparison, present-day massive 
ellipticals galaxies within a $\sim$1 kpc radius are less dense by a factor $\sim$ seven than our 
distant compact massive 
galaxies. On the other hand, in the outer region reachable with our present observations (5 kpc $<R<$10 kpc) 
modern massive elliptical galaxies are denser than the high$-z$ compact massive galaxies by a factor 
of $\sim$ two.

\begin{table*}
 \centering
 
  \caption{Mean stellar surface mass density at different galaxy physical radii}

  \begin{tabular}{cccccc}
  \hline
Galaxy & Type  & $\rho$($R<$1 kpc)   &  $\rho$(1$<R<$3 kpc)  & $\rho$(3$<R<$5 kpc) & $\rho$(5$<R<$10 kpc)   \\
  ID   &       & 10$^9$\solm/kpc$^2$ & 10$^9$\solm/kpc$^2$  & 10$^9$ \solm/kpc$^2$ &  10$^9$\solm/kpc$^2$  \\

\hline
Disk-like (n$<$2.5) \\
\hline

12008481 & normal  & 9.2  & 3.8  &  1.5 & 0.4  \\
22101272 & compact & 15.0 & 6.3  &  2.0 & 0.3 \\

SDSS     & normal  & 3.8 & 1.8   &  1.0 & 0.4 \\

\hline
Spheroid-like (n$>$2.5) \\
\hline

12016582 & normal  &  11.6  &  2.5 &  0.9  & 0.3   \\
22022016 & normal  &  26.6  &  3.21 &  0.9  & 0.3   \\
22026207 & compact & 39.4   &  5.6 &  1.2  & 0.2   \\
22030788 & compact & 42.0   &  5.0 &  1.1  & 0.2  \\
22035487 & normal  & 24.0   &  3.8 &  1.1  & 0.3  \\
22044691 & compact & 57.4   &  4.3 &  0.7  & 0.1  \\
SDSS     & normal  & 6.7    &  1.9 &  0.8  & 0.3  \\

\hline
\label{data}
\end{tabular}
\end{table*}

Our results confirm and expand on previous claims that compact galaxies are growing inside-out,
 adding new  mass in the outer wings as time progresses. We are however also able to show, 
due to our high spatial resolution imaging, that the inner 
regions of the compact galaxies also evolve with time, decreasing their stellar
mass density by a significant factor (see also, Stockton et al. 2010). The amount of this 
evolution will depend on which type of galaxies our compact objects end up evolving into. 
For example, if our compact galaxies become the cores of the most massive, and larger
S\'ersic index $n$, present-day galaxies, the decrease in the inner mean density could 
be as small a factor $\sim$ two.

\subsection{Structural Distortions}

Five out of eight of our objects display a smooth structure, without much substructure, or 
evidence for any dynamical activity, such as interactions, accretion, or mergers with other 
galaxies (Figure~1). 
In particular, three of these smooth galaxies have small sizes with r$_e<$3 kpc. Compared to equal mass 
present-day galaxies these objects are very compact. However, a surprising result from our
analysis is that our sample galaxies are not all smooth and structure-less systems.  Three of our 
eight systems display such dynamical activity (12008481, 22035487 and 22101272), which is  
further evidence that galaxies of very high stellar mass at $z>1$ are potentially still 
forming and evolving through mergers at this epoch (e.g., Bluck et al. 2009). The
features we see are  subtle however, and could be easily missed when viewed at bluer 
wavelengths, or shallower NIR data.  The fact that these galaxies show signs of tidal disturbances, 
implies that at some time before we observe them, these systems likely underwent a merger or 
accretion episode. Interestingly, two of our distorted systems (12008481 and 22101272) have 
a disk-like structure according to their surface brightness profiles, with a lower value of the 
S\'ersic index $n$. The fraction of massive galaxies with lower S\'ersic index $n$ at high$-z$ is 
found to increase with redshift (see e.g. Trujillo et al. 2007; Buitrago et al. 2008; van 
Dokkum et al. 2009). This result suggests that the most massive galaxies are likely  
formed in a disk-like structure that is altered to evolve into present-day
elliptical-like profiles.

The characteristics of these distortions are listed below.  We find that the system 
POWIR 12008481 at $z = 1.23$  displays clear evidence for a tidal distortion over 10 kpc.  
This feature resembles in the local universe tidal plumes produced when galaxies 
undergo a merger with another galaxy.  The fate of this galaxy is unknown, although  it is 
unlikely  that this system becomes a modern massive disk  given its high stellar mass, and 
the few suitable massive disk galaxies found in the local universe (e.g., Conselice 2006). It 
is more likely that after the merger/accretion event this galaxy evolves into a normal size 
massive present-day elliptical. Another peculiar disk-like system is POWIR 22101272 which 
shows a slight distorted and elongated morphology and may also have tidal debris or an interacting 
companion in the direction of its major axis. POWIR 22044691 also has a possible companion 
galaxy which is 12 kpc away, but there is no evidence for an interaction. Finally, another object 
which shows similar outer light asymmetries is the galaxy POWIR 22035487. These asymmetries 
are more visible, and easier to see, after subtracting the best fit light profile from this system. 
This galaxy is however clearly asymmetric in its outer portions.  This is our second largest galaxy 
in the sample, and has a S\'ersic index $n= 4.6$, suggesting that this could be the final product 
of a merger.

Overall, we have a minimum of three systems with readily visible disturbances, giving us an 
active fraction of 0.38$\pm$0.20, which is much higher than the major merger fraction of massive 
galaxies at this epoch (e.g., Conselice et al. 2003, 2009; Bluck et al. 2009).  If we assume that 
these disturbances last on the order of a Gyr, then the duty cycle of this activity, and the 
observed fraction which are disturbed gives a galaxy merger rate (see Bluck et al. 2009) of
$\Gamma = 2.6$ Gyr,  implying that a typical galaxy will undergo one of these disturbances roughly 
every 2.5 Gyr.   Assuming no decline with lower redshifts, this gives us a upper limit, between 
$z = 2$ and today, of five merger events. Since the major merger rate declines with
redshift, the total number of minor mergers is likely lower than this. This is much higher 
than the major merger rate, and suggests that these events are produced through minor 
mergers, or some type of accretion activity.  Future observations will reveal whether this light
originates from newly formed stars such as through cold gas accretion, or through minor merger 
events.

\section{Summary}

We present new NIRI-Altair/LGS adaptive optics observations of eight massive, 
M$_{*}$$\sim$4$\times$\mass, galaxies at $1<z<2$, for which we have measured structural 
parameters and effective radii.  We confirm through using the observed
K-band profiles that some of these systems are indeed very compact, with sizes 
$\lesssim$2 kpc, a factor of $\sim$ seven smaller than similar mass galaxies 
in the local universe. Our observations combine a unique depth and resolution to
explore the stellar mass density profiles of massive galaxies at high$-z$ on an individual 
galaxy basis. We show for the first time that these galaxies are not only growing 
inside-out, but that there is evolution in their inner regions, such that the inner
density declines with lower redshifts. The amount of evolution within our compact galaxies 
at high-z strongly
depends upon which type of galaxies these objects evolve into in the local universe.

In addition, we show that at these redshifts, our systems display evidence for mergers 
or accretion events which are potentially driving the evolution of these systems, particularly 
their sizes and their overall morphology in terms of their surface brightness profiles. 
We find that the systems with a `disk-like' 
morphology with $n \sim 1$ have peculiar structures, suggesting that these systems
are currently relaxing from a merger or accretion event. It remains possible that their 
outer light will eventually accrete back onto each galaxy, thereby increasing its outer 
light, and thus increasing its measured size.

\section*{Acknowledgments}
We would like to thank the anonymous referee for the useful comments and suggestions
that helped to improve the quality of the article. These results in this paper are based on observations 
obtained at the Gemini Observatory, which is operated  by the Association of Universities for 
Research in Astronomy, Inc., under a cooperative agreement with the NSF on behalf of the Gemini partnership: 
the National Science Foundation (United States), the Science and Technology 
Facilities Council (United Kingdom), the National Research Council (Canada), 
CONICYT (Chile), the Australian Research Council (Australia), Minist\'erio da 
Ci\^encia e Tecnologia (Brazil) and Ministerio de Ciencia, Tecnolog\'{\i}a e 
Innovaci\'on Productiva  (Argentina). The observations were carried out as part 
of programs GN-2008A-Q-53 and GN-2009A-Q-33.

\appendix

\label{lastpage}

\end{document}